\documentclass[useAMS,usenatbib]{mn2e}
\usepackage{graphicx}
\usepackage{url}
\pdfoutput=1 
\title[The supermassive black hole in NGC\,4486a]{The supermassive black hole in NGC\,4486a detected with SINFONI at the VLT\thanks{Based on observations at the European
Southern Observatory VLT (075.B-0236)}\thanks{Based on
observations made with the Advanced Camera for Surveys on board the
NASA/ESA {\it Hubble Space Telescope} (GO Proposals 9401), obtained
from the ESO/ST-ECF Science Archive Facility.}}

\author[N. Nowak et al.]{N. Nowak$^{1}$\thanks{E-mail:
    nnowak@mpe.mpg.de}, R.~P. Saglia$^{1}$, J. Thomas$^{2}$, R. Bender$^{1, 2}$, M. Pannella$^{1}$, \newauthor K. Gebhardt$^{3}$, R.~I. Davies$^{1}$\\
$^{1}$Max-Planck-Institut f\"{u}r extraterrestrische Physik, Giessenbachstrasse, 85741 Garching, Germany\\
$^{2}$Universit\"{a}tssternwarte, Scheinerstrasse 1, 81679 Munich, Germany\\
$^{3}$Astronomy Department, University of Texas, Austin, TX~78723}

\begin{document}


\pagerange{\pageref{firstpage}--\pageref{lastpage}} \pubyear{2007}

\maketitle

\label{firstpage}

\begin{abstract}
The near-infrared integral field spectrograph SINFONI at the ESO VLT
opens a new window for the study of central supermassive black
holes. With a near-IR spatial resolution similar to HST optical and
the ability to penetrate dust it provides the possibility to explore
the low-mass end of the $M_\bullet$-$\sigma$ relation
($\sigma<120$~km~s$^{-1}$) where so far very few black hole masses
were measured with stellar dynamics.  With SINFONI we observed the
central region of the low-luminosity elliptical galaxy NGC\,4486a at a
spatial resolution of $\approx0.1\arcsec$ in the $K$ band. The stellar
kinematics was measured with a maximum penalised likelihood method
considering the region around the CO absorption band heads. We
determined a black hole mass of
$M_{\bullet}=(1.25^{+0.75}_{-0.79})\times10^7~\mathrm{M_{\odot}}$
(90~\% C.L.) using the Schwarzschild orbit superposition method
including the full 2-dimensional spatial information. This mass agrees
with the predictions of the $M_\bullet$-$\sigma$ relation,
strengthening its validity at the lower $\sigma$ end.
\end{abstract}


\section{Introduction}

Studies of the dynamics of stars and gas in the nuclei of nearby
galaxies during the last few years have established that all galaxies
with a massive bulge component contain a central supermassive black
hole (SMBH; \citealt{Richstone-98,Bender-03}). Masses of these SMBHs ($M_{\bullet}$)
are well correlated with the bulge luminosity or mass, respectively,
and with the bulge velocity dispersion $\sigma$
\citep{Gebhardt-00a,Ferrarese-00,Kormendy-95}. There are, however,
still a lot of open questions in conjunction with the $M_\bullet$-$\sigma$
correlation, among them the exact slope, its universality and the
underlying physics. A more precise knowledge of the behaviour of the
$M_\bullet$-$\sigma$ relation would help to constrain theoretical
models of bulge formation and black hole growth
(e.g. \citealt{Silk-98,Burkert-01,Haehnelt-00} and others).

In inactive galaxies, the evidence for the existence of black holes and their
masses comes from gravitational effects on the dynamics of stars
inside the black hole's sphere of influence. Since the radius of the
sphere of influence scales with the black hole mass $M_\bullet$, high
resolution observations are needed to detect SMBHs in the low-mass
regime, even for the most nearby galaxies. Further difficulties arise from
the presence of dust in many of these galaxies, particularly
in discs, which enforces observations in the infrared.  Up to now the
low-mass regime ($\sigma\la120$~km~s$^{-1}$) is sparsely sampled with only
three dynamical black hole mass measurements
(Milky Way, \citealt{Schoedel-02}; M32, \citealt{Verolme-02}; NGC\,7457 \citealt{Gebhardt-03}) and some upper limits
(e.g. \citealt{Gebhardt-01,Valluri-05}). Since the near-infrared
integral-field spectrograph SINFONI became operational
\citep{Eisenhauer-sinfoni,Bonnet-sinfoni}, it is now possible to detect low-mass black
holes in dust-obscured galaxies at a spatial resolution close to that
of HST.

NGC\,4486a is a low-luminosity elliptical galaxy in the Virgo cluster,
close to M87. It contains an almost edge-on nuclear disc of stars and
dust \citep{Kormendy-4486a}. The bright star $\sim2.5\arcsec$ away
from the centre makes it impossible to obtain undisturbed spectra with
conventional ground-based longslit spectroscopy. However, it is one of
the extremely rare cases, where an inactive, low-luminosity galaxy can
be observed at diffraction limited resolution using adaptive optics
with a natural guide star (NGS). This feature made NGC\,4486a
one of the most attractive targets during the years between the
commissioning of SINFONI and similar instruments and the installation
of laser guide stars (LGS). NGC\,4486a is the first of our sample
of galaxies observed or planned to be observed using near-infrared
integral-field spectroscopy with the goal to tighten the slope of the
$M_\bullet$-$\sigma$ relation in the low-$\sigma$ regime
($\la120~\mathrm{km~s^{-1}}$) and for pseudobulge galaxies.

This paper is structured as follows: In \S 2 we present the
observations and the data reduction.  The derivation of the kinematics
and the photometry is described in \S 3 and \S 4 and the dynamical
modelling procedure is explained in \S 5. In \S 6 the
results are presented and discussed and conclusions are drawn.

\section{Observations and Data Reduction}

NGC\,4486a was observed on April 5 and 6, 2005, as part of guaranteed
time observations with SINFONI \citep{Eisenhauer-sinfoni,Bonnet-sinfoni} at the
VLT UT4. SINFONI consists of the integral-field spectrograph SPIFFI
(spectrometer for infrared faint field imaging;
\citealt{Eisenhauer-spiffi}) and the curvature-sensing
adaptive optics (AO) module MACAO \citep{Bonnet-macao}. We used the
$K$ band grating (1.95--2.45~$\mathrm{\mu}$m) and the $3\arcsec\times3\arcsec$
field of view (0.05$\arcsec\times0.1\arcsec$ px$^{-1}$). The bright
($R\approx11$~mag) star located $\sim2.5\arcsec$ southwest of the
nucleus was used for the AO correction. The seeing indicated by the
optical seeing monitor was between 0.6$\arcsec$ and 0.85$\arcsec$,
resulting in a near-infrared seeing better than $\sim0.7\arcsec$,
which could be improved by the AO module to reach a resolution of
$0.1\arcsec$ ($\sim25$\% Strehl). For the chosen configuration
SINFONI delivers a nominal FWHM spectral resolution of $R\approx4400$.
In total 14 on-source and 7 sky exposures of 600s each were taken in
series of ``object--sky--object'' cycles, dithered by up to
0.2$\arcsec$. During the last observation block the visual seeing
suddenly increased above 1.0$\arcsec$. Therefore, to keep the spatial
resolution as high as possible, only exposures with a seeing
$\la0.85\arcsec$ were considered for the derivation of the
kinematics. In order to derive the point-spread function (PSF) an
exposure of the AO star was taken regularly.
\begin{figure}
\centering
\includegraphics[width=.85\linewidth,keepaspectratio]{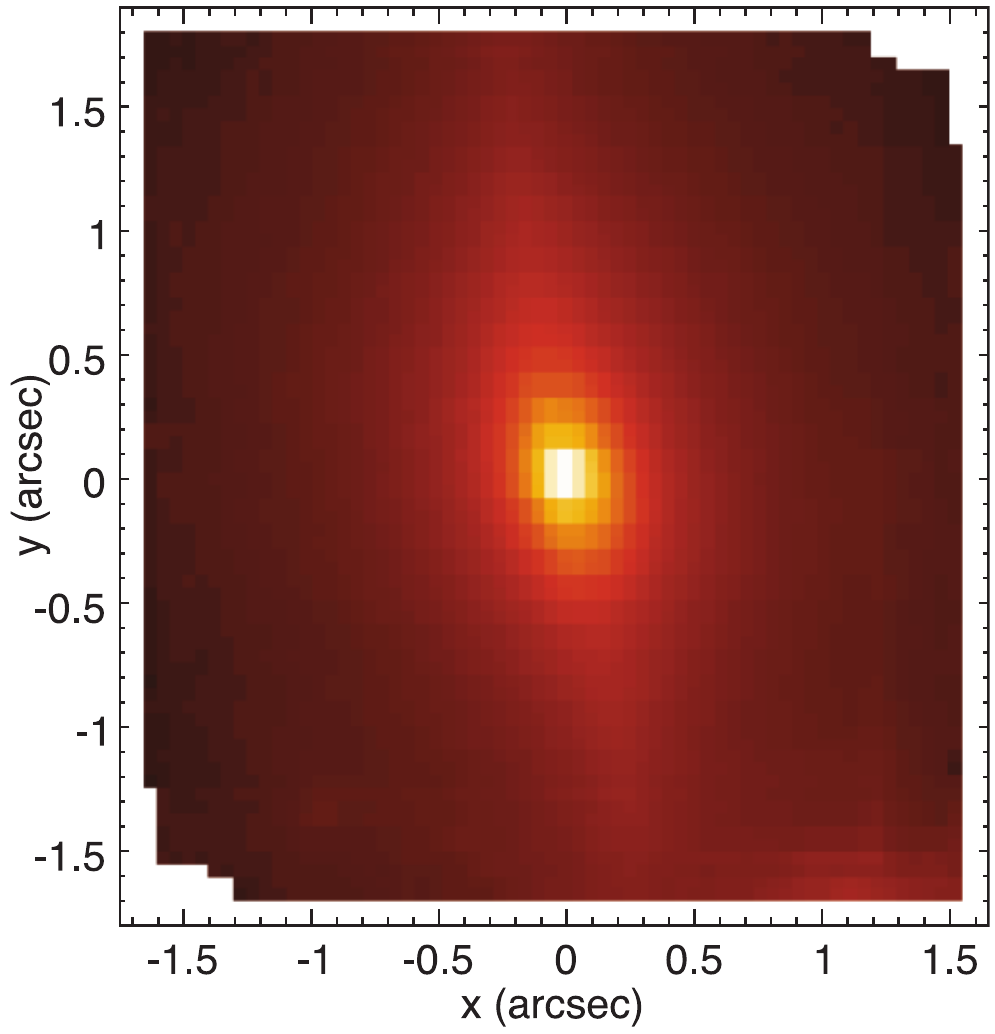}
\caption{SINFONI image of NGC\,4486a.}\label{fig:1}
\end{figure}

The SINFONI data reduction package {\sc spred}
\citep{Schreiber-spred,Abuter-spred} was used to reduce the data. It
includes all common reduction steps necessary for near-infrared data
plus routines to reconstruct the three-dimensional data cubes. After
subtracting the sky frames from the object frames, the data were
flatfielded, corrected for bad pixels, for distortion and wavelength
calibrated using a Ne/Ar lamp frame. The wavelength calibration was
corrected using night-sky lines if necessary. Then the
three-dimensional data cubes were reconstructed and corrected for
atmospheric absorption using the B9.5V star Hip~059503. As a final
step all data cubes were averaged together to produce the final data
cube. Fig. \ref{fig:1} shows the resulting SINFONI image (collapsed
cube) of NGC\,4486a.  The data of the telluric and the PSF stars were
reduced likewise.

\begin{figure}
\centering
\includegraphics[width=\linewidth,keepaspectratio]{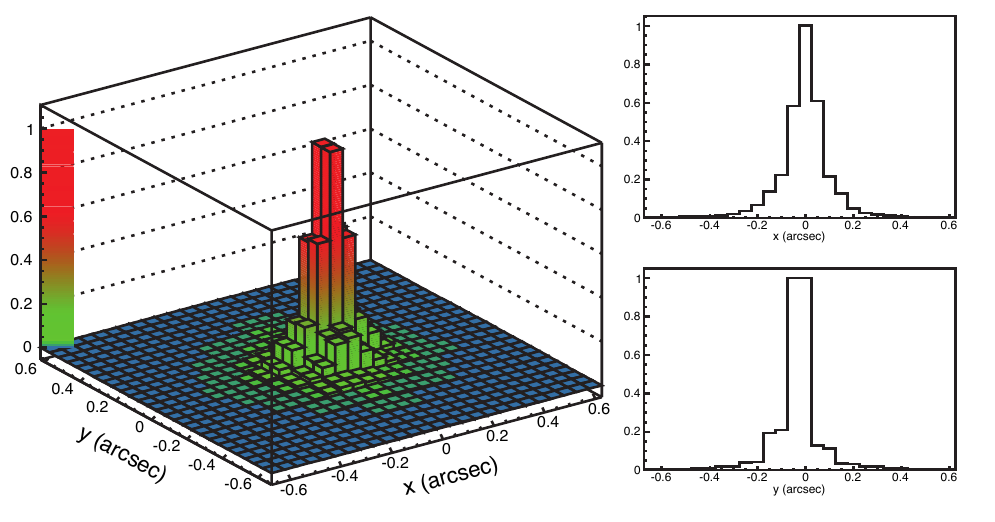}
\caption{Left: The two-dimensional PSF. Right: x- and y-profiles of the PSF.}\label{fig:2}
\end{figure}

As PSF we take the combined and normalised image of the PSF star
exposures (see Fig. \ref{fig:2}). Its core can be reasonably well
fitted by a Gaussian profile with a FWHM of $\sim100$~mas
(7.7~pc). The diameter of the sphere of influence of the black hole can
be roughly estimated to $130$~mas using the
$M_\bullet$-$\sigma$ relation of \citet{Tremaine-02} and is therefore
resolved.

\section{Kinematics}

The kinematic information was extracted using the maximum penalised
likelihood (MPL) technique of \citet{Gebhardt-00c}, which obtains
non-parametric line-of-sight velocity distributions (LOSVDs). As
kinematic template stars we use six K0 to M0 stars which were observed
during commissioning and our GTO observations in 2005 with SINFONI
using the same configuration as for NGC\,4486a. Both galaxy and
template spectra were continuum-normalised. An initial binned velocity
profile is convolved with a linear combination of the template spectra
and the residuals of the resulting spectrum to the observed galaxy
spectrum are calculated. The velocity profile is then changed
successively and the weights of the templates are adjusted in order to
optimise the fit to the observed spectrum by minimizing the function
$\chi^2_\mathrm{p}=\chi^2+\alpha\mathcal{P}$, where $\alpha$ is the
smoothing parameter that determines the level of regularisation, and
the penalty function $\mathcal{P}$ is the integral of the square of
the second derivative of the LOSVD.  We fitted only the first two band
heads CO2--0 and CO3--1. The higher-order band heads are strongly
disturbed by residual atmospheric features. At wavelength
$\lambda<2.29~\mathrm{\mu}$m the absorption lines are weak and cannot
be fitted very well by the templates.

The uncertainties on the velocity profiles were estimated using Monte
Carlo simulations \citep{Gebhardt-00c}. A galaxy spectrum is created
by convolving the template spectrum with the measured LOSVD. Then 100
realisations of that initial galaxy spectrum are created by adding
appropriate Gaussian noise.
The LOSVDs of each realisation are determined and used to specify the
confidence intervals.

In order to test the performance of the method on our SINFONI data and
to find the best fitting parameters we performed Monte Carlo
simulations on a large set of model galaxy spectra. These were created
from stellar template spectra by convolving them with both Gaussian
and non-Gaussian LOSVDs and by adding different amounts of noise. We
found that the reconstructed LOSVDs resemble the input LOSVDs very
well if the smoothing parameter $\alpha$ is chosen adequately
\citep{Merritt-97,Joseph-01}. The best choice of $\alpha$ solely
depends on the signal-to-noise ratio (S/N) of the data.  To maximize
the S/N of the data a binning scheme with 11 radial and 5 angular bins
per quadrant, similar to that used in \citet{Gebhardt-03}, was
chosen. The centres of the angular bins are at latitudes
$\vartheta=5.8^\circ$, $17.6^\circ$, $30.2^\circ$, $45.0^\circ$ and
$71.6^\circ$ from the major to the minor axis. The bins are not
overlapping, but spatial resolution elements at the border between
bins may be divided into parts where each part is counted to a
different bin. The spectra within each bin were averaged with weigths
according to their share in the bin. The radial binning scheme ensures
that an adequate S/N level comparable to that of the central spectrum
(S/N$\approx40$) is maintained at all radii at the cost of spatial
resolution outside the central region.

\begin{figure}
\centering
\includegraphics[width=\linewidth,keepaspectratio]{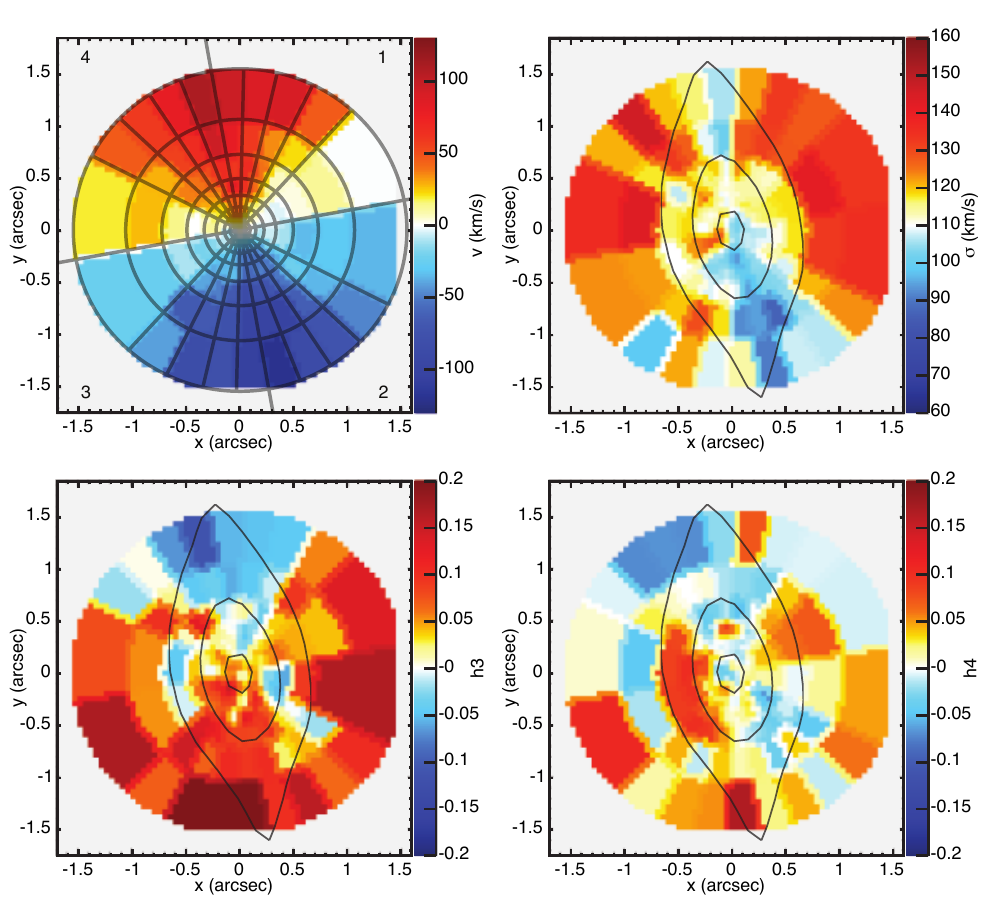}
\caption{Stellar kinematic fields ($v$, $\sigma$, $h_3$, $h_4$) of
NGC\,4486a. The binning scheme and the numbering of the quadrants are
overlaid over the velocity map and the isophotes are overlaid over the
other maps for comparison.}\label{fig:3}
\end{figure}

\begin{figure}
\centering
\includegraphics[width=.9\linewidth,keepaspectratio]{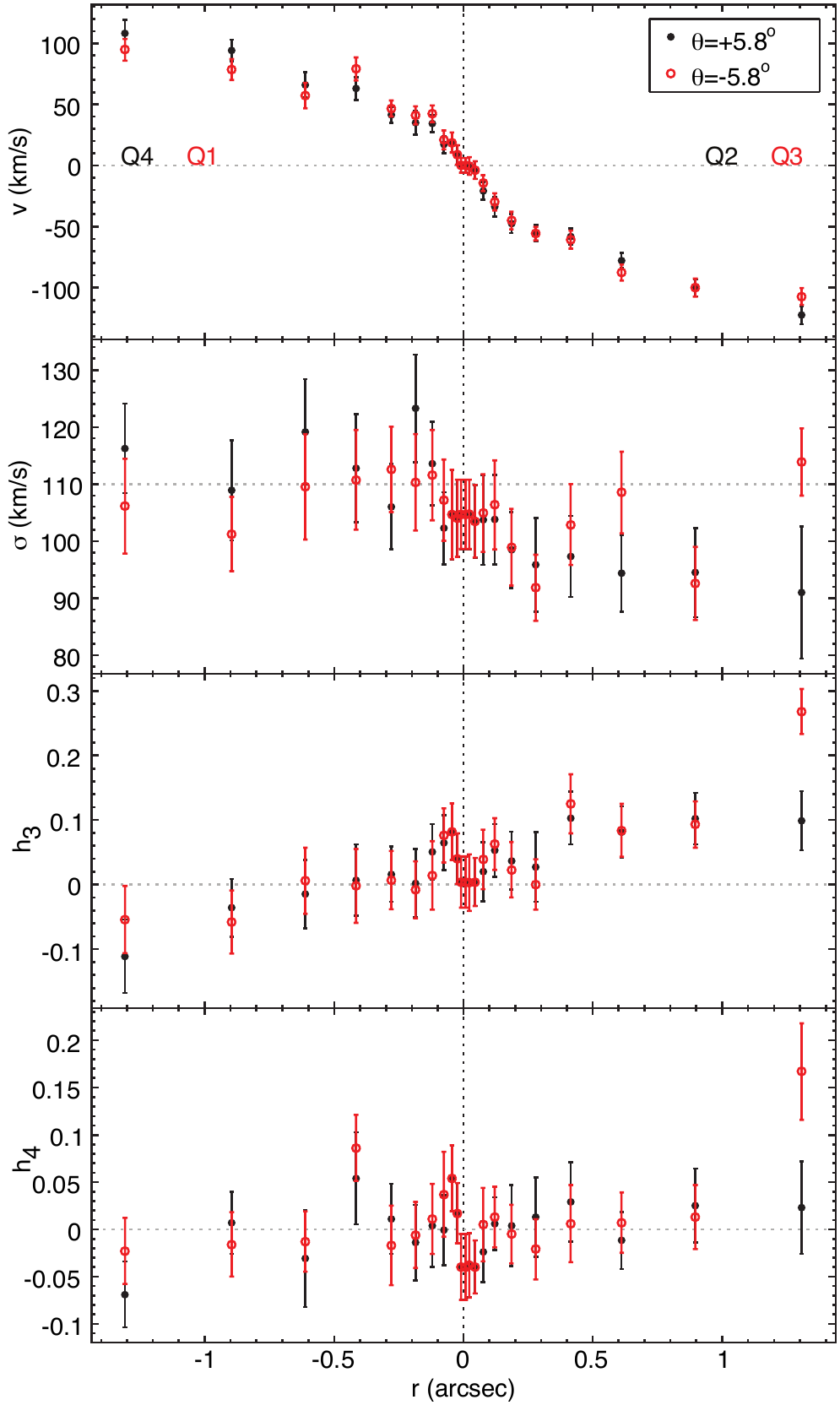}
\caption{Stellar kinematic profiles ($v$, $\sigma$, $h_3$, $h_4$) of
NGC\,4486a along $\theta=\pm5.8^\circ$ of the major axis. The corresponding quadrants crossed by the profiles are marked for easy identification in Fig. \ref{fig:3}.}\label{fig:4}
\end{figure}

The resulting two-dimensional kinematics ($v$, $\sigma$, $h_3$, $h_4$)
is presented in Fig. \ref{fig:3}. It illustrates the superposition of
the kinematics of the two distinct components in NGC\,4486a -- the
disc and the bulge. Whereas the velocity map shows a regular rotation
pattern, the cold stellar disc can be clearly distinguished from the
surrounding hotter bulge in the velocity dispersion map. The velocity
dispersion of the disc is $\approx20-30$~km~s$^{-1}$ smaller than that
of the bulge. Asymmetric and symmetric deviations from a Gaussian
velocity profile are quantified by the higher-order Gauss-Hermite
coefficients $h_3$ and $h_4$
\citep{Gerhard-93,Marel-93}. Fig. \ref{fig:4} shows the kinematic
profiles of NGC\,4486a along the major axis at angles
$\theta=+5.8^\circ$ and $\theta=-5.8^\circ$. The
$+5.8^\circ$-profile agrees very well with the adjacent
$-5.8^\circ$-profile within the error bars. When comparing the
profiles at negative radii with the profiles at positive radii slight
asymmetries can be seen (especially in $\sigma$), but as the errors
are relatively large, deviations from axisymmetry are small.

The star at $2.5\arcsec$ from the centre, which heavily dilutes optical
spectra of NGC\,4486a taken without AO, does not have any significant
effect on the kinematics derived here. Quadrant $2$ is affected most, as
it is located in the direction of the star. The fraction of inshining
light from the star is only about 13\% at $0.7\arcsec$ from the centre
of the galaxy in the direction of the star (the outermost point
covered by the exposures of the PSF star) due to the narrow PSF. In
addition the spectrum of the star shows neither the strong CO
absorption nor other spectral features in that wavelength region.

\section{Imaging}

To derive the black hole mass in NGC\,4486a, it is essential to
determine the gravitational potential made up by the stellar component
by deprojecting the surface brightness distribution. As NGC\,4486a
consists both kinematically and photometrically of two components with
possibly different mass-to-light ratios $\Upsilon$, we deproject bulge
and disc separately.

To decompose the two components, we considered the HST images in the
broad-band F850LP filter, with 2 ACS/WFC pointings of 560 seconds
exposure each. The two dithers have no shift in spatial
coordinates. The data were reduced by the ST-ECF On-The-Fly
Recalibration system, see \url{http://archive.eso.org/archive/hst} for
detailed information.

Moreover, we use the {\sc galfit} package \citep{Peng-02} to fit PSF
convolved analytic profiles to the two-dimensional surface brightness
of the galaxy.  The code determines the best fit by comparing the
convolved models with the science data using a Levenberg-Marquardt
downhill gradient algorithm to minimize the $\chi^2$ of the fit.  The
saturated star close to the galaxy centre has been masked out from the
modelling. The observing strategy, i.e. the adopted no spatial shift
between the two dithers, has allowed us to obtain a careful
description of the PSF by using the
{\sc TinyTim}\footnote{\url{http://www.stsci.edu/software/tinytim/tinytim}}
code.

\begin{figure}
\centering
\includegraphics[width=\linewidth]{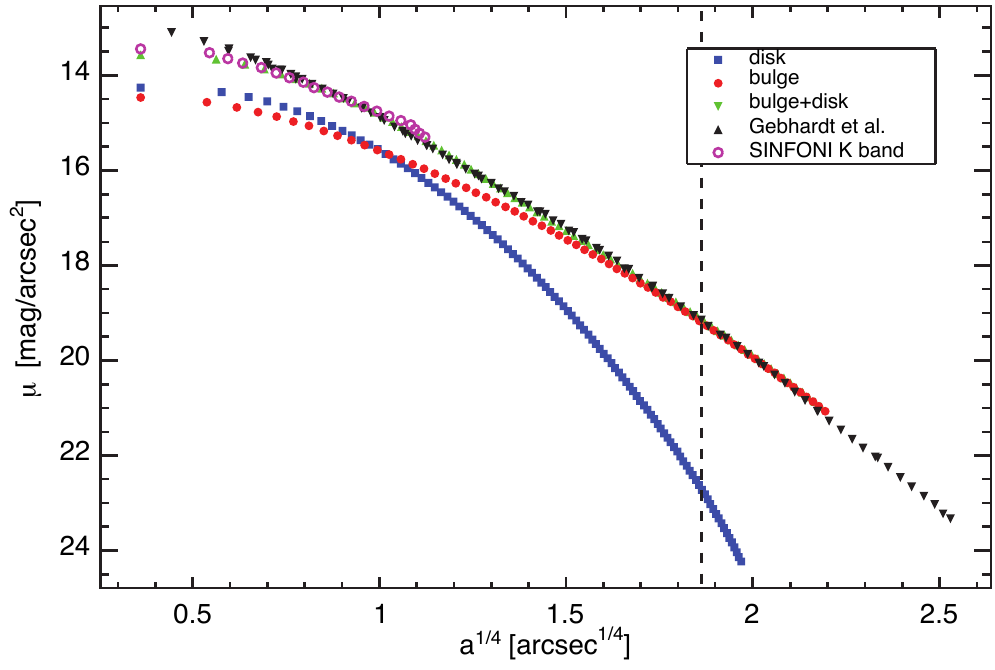}
\caption{Surface brightness profile for NGC\,4486a. At radii left of
the dashed line we used the decomposed bulge and disc profiles for the
deprojection, right of the dashed line we used the profile of Gebhardt
et al. (in prep.) (scaled to ACS $z$-band).}\label{fig:5}
\end{figure}
We modelled the galaxy light with a double \citet{Sersic-68} model
with indices $n=2.19$ for the bulge and $n=1.67$ for the disc. In Fig.
\ref{fig:5} we show the surface brightness profiles of bulge, disc,
bulge+disc, and for comparison the SINFONI surface brightness profile
and the un-decomposed profile of Gebhardt et al. (in prep.), which was
derived from HST and CFHT imaging in several bands. This profile
agrees very well with our combined profile and we use it, scaled to
match our profile, at radii $>11.5\arcsec$ where the bulge strongly
dominates.

Bulge and disc were then deprojected separately using the program of
\citet{Magorrian-99} under the assumption that both components are
edge-on and axisymmetric. The stellar mass density then can be
modelled as in \citet{Davies-06} via
$\rho_{\mathrm{*}}=\Upsilon_{\mathrm{bulge}}\cdot\nu_{\mathrm{bulge}}+\Upsilon_{\mathrm{disc}}\cdot\nu_{\mathrm{disc}}$,
where $\nu$ is the luminosity density obtained from the deprojection
and the ACS $z$-band mass-to-light ratio $\Upsilon$ is
assumed to be constant with radius for both components.

\section{Schwarzschild Modelling}

The mass of the black hole in NGC\,4486a was determined based on the
\citet{Schwarzschild-79} orbit superposition technique, using the code
of \citet{Gebhardt-00c,Gebhardt-03} in the version of
\citet{Thomas-04}. It comprises the usual steps: (1) Calculation of a
potential with a trial black hole of mass $M_\bullet$ and a stellar
mass density $\rho_{\mathrm{*}}$. (2) A representative set of orbits
is run in this potential and an orbit superposition that best matches
the observational constraints is constructed. (3) Repetition of the
first two steps with different values for $\Upsilon_\mathrm{bulge}$,
$\Upsilon_\mathrm{disc}$ and $M_\bullet$ until the eligible parameter
space is systematically sampled. The best-fitting parameters then
follow from a $\chi^2$-analysis.

The models are calculated on the grid with 11 radial and 5 angular bins
per quadrant as described above (cf. Fig. \ref{fig:3}).

Our orbit libraries contain $2\times7000$ orbits. The luminosity
density is a boundary condition and hence exactly reproduced.  The
$11\times5$ LOSVDs are binned into $17$ velocity bins each and then
fitted directly, not the parametrized moments.  We limit the
parameter space for the values of $\Upsilon$ by considering the
population synthesis model of \citet{Maraston-98,Maraston-05}, which
gives us $\Upsilon\la5$ for the $z$-band.

Special care was taken when implementing the
PSF. Due to its special shape with the narrow core and the broad wings
(cf. Fig. \ref{fig:2}) the PSF was not fitted, rather the
two-dimensional image of the star was directly used for convolving our
models.

\begin{table}
\caption{Results obtained for the four quadrants separately with a global mass-to-light ratio (90\% C.L.). The numbering of the quadrants can be inferred from Fig. \ref{fig:3}. \label{tab:1}}
\begin{tabular}{ccc}
\hline\hline
Quadrant & $M_{\bullet}$ ($10^7$\,M$_\odot$) & $\Upsilon$ \\
\hline
1  & $4.0^{+0.7}_{-2.4}$ &  $3.4^{+0.6}_{-0.6}$\\
2  & $1.0^{+1.1}_{-0.2}$ &  $3.6^{+0.2}_{-0.5}$\\
3  & $1.0^{+0.5}_{-0.5}$ &  $3.8^{+0.6}_{-0.1}$\\
4  & $1.5^{+1.0}_{-1.0}$ &  $4.4^{+0.3}_{-0.9}$\\
$1-4$ averaged & $1.25^{+0.3}_{-0.3}$ & $4.0^{+0.1}_{-0.4}$\\
\hline
\end{tabular}
\end{table}

\section{Results}

\begin{figure}
\centering
\includegraphics[width=\linewidth,keepaspectratio]{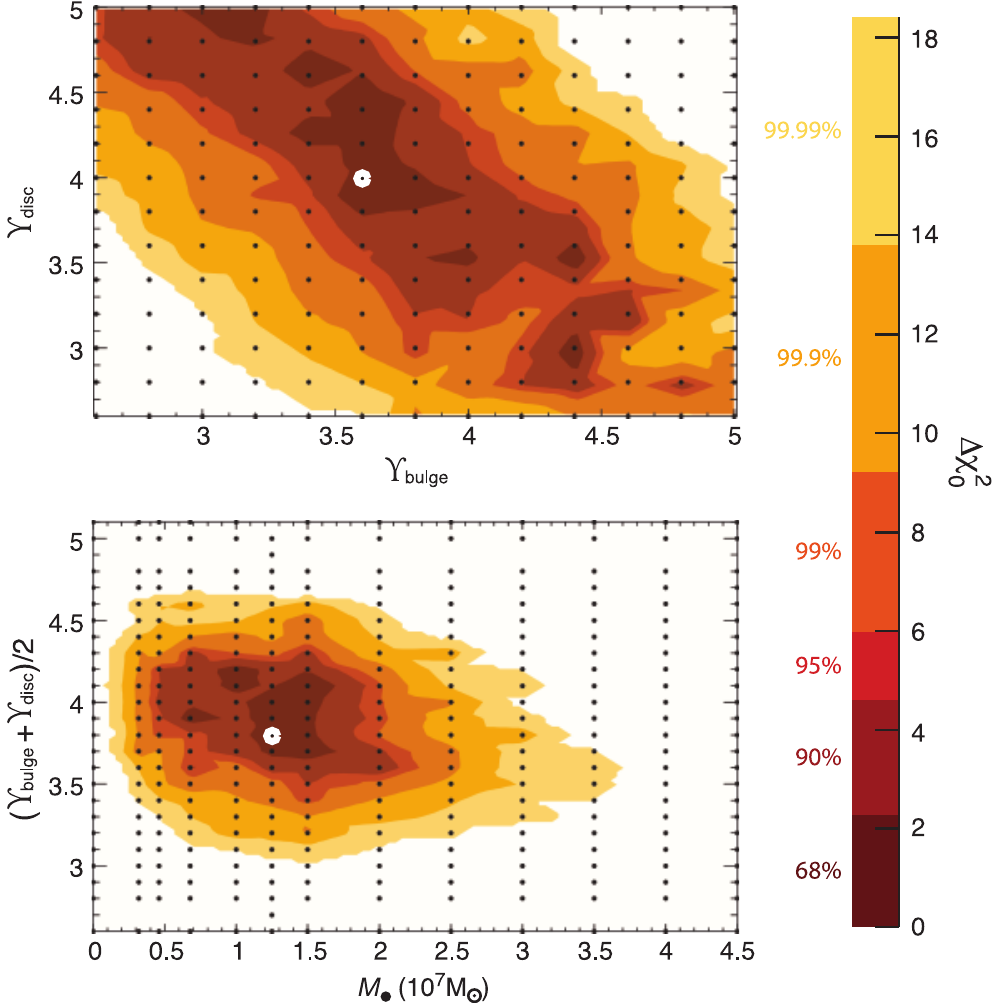}
\caption{$\Delta\chi_0^2=\chi^2-\chi_\mathrm{min}^2$ as a function of 
(top) $\Upsilon_\mathrm{bulge}$ and $\Upsilon_\mathrm{disc}$,
minimized over $M_\bullet$; (bottom) $M_\bullet$ and
$\left(\Upsilon_{\mathrm{bulge}}+\Upsilon_{\mathrm{disc}}\right)/2$.
The black points are the models we calculated and the coloured regions
are the (unsmoothed) confidence intervals for two degrees of
freedom. The best-fitting model is marked with a white
circle.}\label{fig:6}
\end{figure}
A big advantage of integral-field data compared to longslit data is
that we can check the assumption of axisymmetry by comparing the
kinematics of the four quadrants and quantify the effect of possible
deviations by modelling each quadrant separately. We do not find major
differences in the kinematics of the four quadrants
(cf. Figs. \ref{fig:3}, \ref{fig:4}). As it takes a large amount of
computing time to calculate all models with different mass-to-light
ratios for bulge and disc for each quadrant, we used the same
$\Upsilon$ for bulge and disc for the comparison of the quadrants. In
Table \ref{tab:1} the resulting values for $M_\bullet$ and $\Upsilon$
are listed. They show that the four quadrants agree reasonably well
with each other. The only systematically deviant point is the
$M_{\bullet}$ determination in the first quadrant that however has a
large error and therefore is compatible within the 90\% C.L. with the
other three quadrants. Therefore we symmetrised the LOSVDs by taking
for each bin the weight-averaged LOSVDs of the four quadrants and the
corresponding errors. The results of modelling these averaged LOSVDs
are shown in Fig. \ref{fig:6}, where
$\Delta\chi_0^2=\chi^2-\chi_\mathrm{min}^2$ is plotted as a function
of $\left(\Upsilon_{\mathrm{bulge}}+\Upsilon_{\mathrm{disc}}\right)/2$
and $M_\bullet$ with error contours for two degrees of
freedom. $\Upsilon_\mathrm{bulge}$ and $\Upsilon_\mathrm{disc}$
anticorrelate such that their sum is approximately constant, as shown
in the upper part of Fig. \ref{fig:6}.  A black hole mass of
$(1.25^{+0.75}_{-0.79})\times10^7~\mathrm{M_{\sun}}$ (90\% C.L.) can
be fitted with $\Upsilon_\mathrm{disc}\approx2.8\dots5.2$ and
$\Upsilon_\mathrm{bulge}\approx2.8\dots4.8$. This result agrees within
90\% confidence limit with the results of all quadrants shown in Table
\ref{tab:1}. The best-fitting model, obtained with minimal
regularisation, is marked with a white circle and has a black hole
mass $M_\bullet=1.25\times10^7\mathrm{M_{\sun}}$ and mass-to-light
ratios $\Upsilon_\mathrm{disc}=4.0$ and $\Upsilon_\mathrm{bulge}=3.6$.
The difference in $\chi^2$ to the best-fitting model without black
hole is $24.14$ which corresponds to $4.5\sigma$. The total $\chi^2$
values for the models are around $300$. Together with the number of
observables ($11$ radial bins $\times$ $5$ angular bins $\times$ $17$
velocity bins) this gives a reduced $\chi^2$ of $\approx0.3$. Note,
however, that the number of observables is in reality smaller due to
the smoothing \citep{Gebhardt-00c}.

The dynamical mass-to-light ratios of disc and bulge agree with an old
and metal-rich stellar population
\citep{Maraston-98,Maraston-05}. $\Upsilon_\mathrm{disc}$ tends to be
larger than $\Upsilon_\mathrm{bulge}$ which is probably due to the
presence of dust in the disc. To estimate the effect of the dust on
the mass-to-light ratio of the disc we are using the model of
\citet{Pierini-04}. For the HST-F850LP filter and an assumed typical
optical depth $\tau\la0.5$ we obtain an attenuation of
$A_\lambda\la0.36$~mag. This translates the best-fitting
$\Upsilon_{\mathrm{disc}}=4.0$ to a significantly smaller
dust-corrected value of $\ga2.9$. Following the models of
\citet{Maraston-98,Maraston-05} this is in good agreement with an
estimated $\ga2$~Gyr younger disc \citep{Kormendy-4486a}.

\begin{figure*}
\centering
\includegraphics[width=\linewidth]{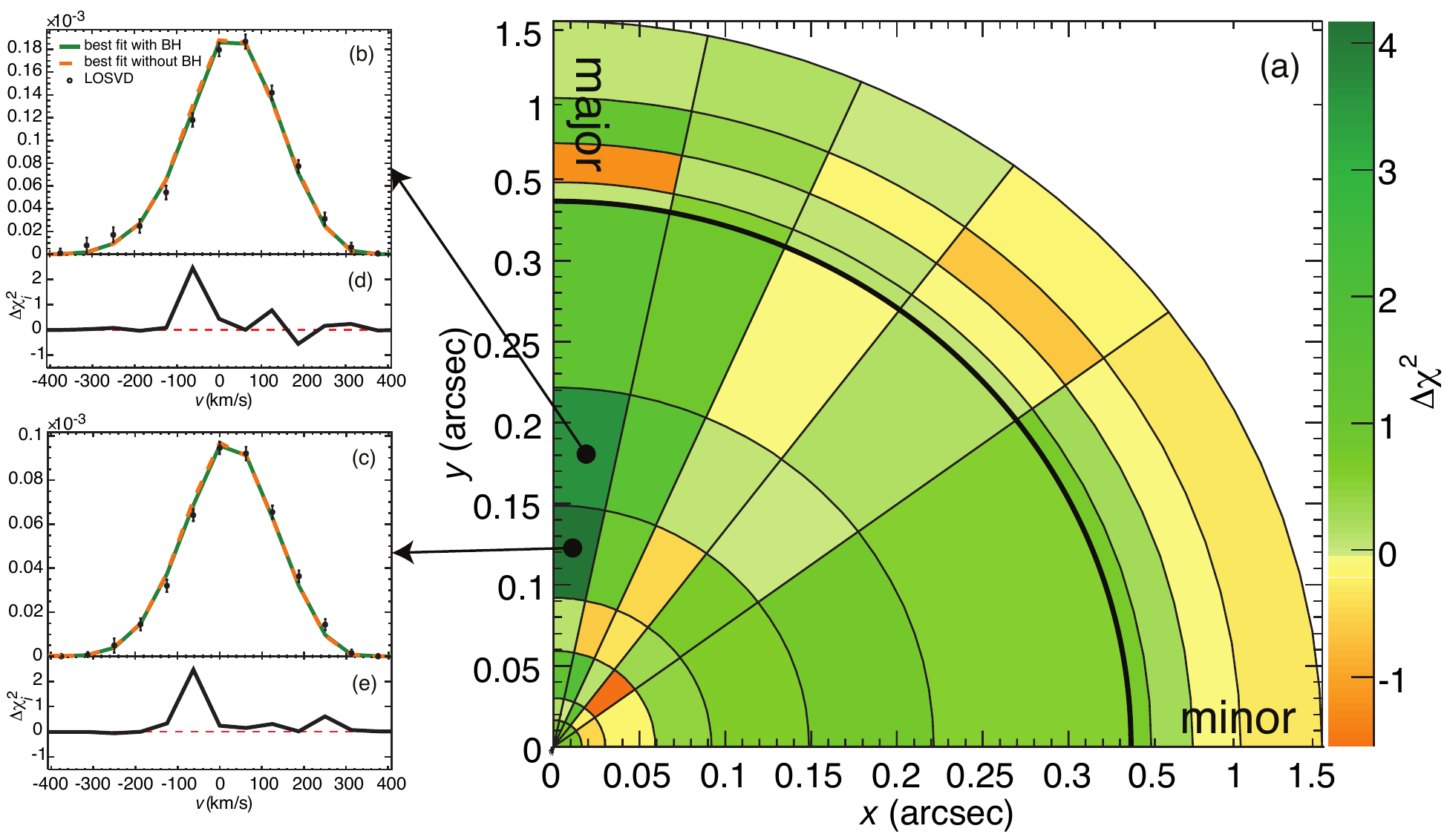}
\caption{(a) $\chi^2$ difference between the best-fitting model
without black hole and the best-fitting model with black hole
($\Delta\chi^2=\sum_i{\Delta\chi_i^2}=\sum_{i=1}^{17}{\left(\chi^2_{i,\mathrm{no
BH}}-\chi^2_{i,\mathrm{BH}}\right)}$ over all 17 velocity bins) for
all LOSVDs of the averaged quadrant. Bins where the model with black
hole fits the LOSVD better are plotted in green, the others in
orange. The part outside $\approx2.5$ spheres of influence is plotted
with a different scale than the inner part, since in the outer region
the dynamical effect of the black hole (and therefore the difference
between the two models) is negligible. (b,c) For the radii with the largest
positive $\chi^2$ difference in (a) the LOSVD (open circles with
error bars, normalised as in \citealt{Gebhardt-00c}) and both fits
(with black hole, full green line, and without black hole, dashed
orange line) are shown with the corresponding $\Delta\chi_i^2$ plotted
below (d,e).}\label{fig:7}
\end{figure*}

The significance of the result is illustrated in Fig. \ref{fig:7}. It
shows the $\chi^2$ difference between the best-fitting model without
black hole and the best-fitting model with black hole
($\Delta\chi^2=\sum_{i=1}^{17}{\left(\chi^2_{i,\mathrm{no
BH}}-\chi^2_{i,\mathrm{BH}}\right)}$ over all 17 velocity bins) for
all LOSVDs of the averaged quadrant. The part outside $\approx2.5$
spheres of influence, where the dynamical effect of the black hole is
negligible, is displayed in a compressed way in order to emphasize the
important inner part.  For $75\%$ of all bins the model with black
hole produces a fit to the LOSVD better than the model without black
hole. The signature of the black hole is imprinted mainly along the
major axis, where the largest positive $\Delta\chi^2$ are found at
radii $r\approx0.09\arcsec\dots0.22\arcsec$. At radii
$r\la0.09\arcsec$ the differences between both models are smaller due
to the effect of the PSF. For the radii with the
largest $\Delta\chi^2$ along the major axis the LOSVD and the fits
with and without black hole are shown in the left part of
Fig. \ref{fig:7} together with the corresponding $\Delta\chi^2_i$ as a
function of the line-of-sight velocity. The differences between the
two fits are relatively small in absolute terms. However, the model
without black hole has more stars on the low-velocity wing at a
$\sim1.5\sigma$ level, failing to match fully the measured slightly
higher mean velocity of the galaxy. Future observations with a higher
spatial resolution should be able to probe this difference more
clearly.

The total stellar mass within $1$ sphere of influence, where the
imprint of the black hole is strongest, is
$M_*=9.84\times10^6~\mathrm{M_{\odot}}$. If the additional mass of
$M_\bullet=1.25\times10^7~\mathrm{M_{\odot}}$ was solely composed of
stars, this would increase the mass-to-light ratio to
$\Upsilon_\mathrm{disc}\approx9.1$ ($6.6$ if we take into account the
dust-absorption), a region which is excluded by stellar population
models or at least requires unrealistic high stellar ages.

The models with black hole become tangentially anisotropic in the
centre, while the models without black hole are close to isotropic.

Our result is in good agreement with the
prediction of the $M_\bullet$-$\sigma$ relation
($1.26^{+0.49}_{-0.35}\times10^7~\mathrm{M_\odot}$ using the result of
\citealt{Tremaine-02}) and strengthens it in the
low-$\sigma$ regime ($\la120$~km~s$^{-1}$), where, besides several upper
limits, up to now only three black hole masses were measured with
stellar kinematics (Milky Way, \citealt{Schoedel-02}; M32,
\citealt{Verolme-02}; NGC\,7457, \citealt{Gebhardt-03}).

NGC\,4486a is only the first object in our sample of low-mass
galaxies under investigation. 
The laser guide star, which is presently being installed at the VLT
UT4, makes observations of a large number of appropriate galaxies now
possible. Therefore we plan to further explore this region of the
$M_\bullet$-$\sigma$ relation by observing more galaxies with velocity
dispersions between the resolution limit of SINFONI ($\sim30$~km~s$^{-1}$)
and $\sim120$~km~s$^{-1}$.

\section*{Acknowledgments}

We are grateful to Frank Eisenhauer and Stefan Gillessen for
assistance in using the SINFONI instrument and the reduction software
{\sc spred}. Furthermore we thank David Fisher for providing us the
surface brightness profile which we used at large radii.

\bibliographystyle{mn2e}
\bibliography{bibliography}




\bsp

\label{lastpage}

\end{document}